\documentclass[12pt]{article}
\usepackage{amsmath}
\usepackage{amsfonts}
\usepackage{amsthm}
\usepackage{amssymb}
\usepackage{bbold}
\DeclareSymbolFontAlphabet{\amsmathbb}{AMSb}
\newtheorem*{theorem}{Theorem}
\newtheorem*{lemma}{Fact}

\newcommand{\basisOne}[1]{\{#1\}}
\newcommand{\basis}[2]{\{#1,#2\}}

\title{Forces on a Clifford bundle}
\author{jason hanson}

\begin{document}
\maketitle

\begin{abstract}
In \cite{me}, the Clifford bundle over spacetime was used as a geometric framework for obtaining coupled Dirac and Einstein equations.  Other forces may be incorporated using minimal coupling.  Here the fundamental forces that are allowed within this framework are explicitly enumerated.
\end{abstract}

%%%%%%%%%%%%%%%%%%%%%%%%%%%%%%%%%%%%%%%%%%%%%%%%%%%%%%%%%%%%%%%%%
\section{Introduction}

In a previous article \cite{me}, we used the Clifford bundle ${\it Cl}_\ast M$ over a spacetime manifold $M$ as a geometric framework for incorporating both the Dirac and Einstein equations.  We also indicated how other forces can be introduced using minimal coupling.  That is, by using a connection on the Clifford bundle of the form $\nabla_\alpha=\partial_\alpha+\hat\Gamma_\alpha+\theta_\alpha$.  Here $\hat\Gamma_\alpha$ is the metric--compatible connection on $M$ extended to ${\it Cl}_\ast M$, and $\theta_\alpha$ is a collection of four $16\times 16$ matrices that encode the force.  These matrices must satisfy certain constraints in order for the variational principle to yield the Dirac equation $\gamma^\alpha\nabla_\alpha\psi=m\psi$.  Such a collection $\theta=\{\theta_0,\theta_1,\theta_2,\theta_3\}$ is a tensor, which we call a {\em force tensor}.

In this article, we enumerate the distinct types of forces that are allowed by minimal coupling.  This is done by considering the action of the Lorentz group on the space of all possible force tensors $V$.  Under this action, $V$ is a representation of the Lorentz group.  And as such, it can be decomposed into irreducible subrepresentations.  Each irreducible subrepresentation corresponds to a distinct force.

{\it Article summary.} In the remainder of this section, we review the relevant constructions used in geometric framework introduced in \cite{me}, as well as the necessary constraints on a force tensor $\theta$.  In section \ref{sec:space}, we find a basis for the space of all possible force tensors $V$, and in section \ref{sec:irred} we write down the irreducible subspaces.  Each irreducible subspace is identified with a type of fundamental force field on spacetime $M$: scalar, vector, anti--symmetric tensor, symmetric tensor, and Fierz tensor.  In section \ref{sec:force}, we use the curvature of the connection $\nabla_\alpha$ to obtain field equations for each of the fundamental forces.

%::::::::::::::::::::::::::::::::::::::::::::::::::::::::::::::::
\subsection{Geometric framework}

Let $M$ be a Lorentz manifold with metric $g$.  Let $x=x^\alpha$, with $\alpha=0,1,2,3$, be local coordinates for $M$, and let ${\bf e}_\alpha\doteq\partial/\partial x^\alpha$ denote the corresponding basis vectors of the tangent bundle of $M$ at $x$.  In this basis, $g_{\alpha\beta}$ denotes the components of $g$, and $g^{\alpha\beta}$ denotes the components of $g^{-1}$.

The Clifford bundle ${\it Cl}_\ast M$ is formed by taking the Clifford algebra of each fiber of the tangent bundle.  That is, we enforce the algebra relation ${\bf e}_\alpha{\bf e}_\beta+{\bf e}_\beta{\bf e}_\alpha=2g_{\alpha\beta}$ on each fiber.  We may choose
$${\bf e}_\emptyset\doteq 1,\,
  {\bf e}_0,\,
  {\bf e}_1,\,
  {\bf e}_2,\,
  {\bf e}_3,\,
  {\bf e}_{01},\,
  {\bf e}_{02},\,
  {\bf e}_{03},\,
  {\bf e}_{12},\,
  {\bf e}_{13},\,
  {\bf e}_{23},\,
  {\bf e}_{012},\,
  {\bf e}_{013},\,
  {\bf e}_{023},\,
  {\bf e}_{123},\,
  {\bf e}_{0123}
$$
as basis vectors for the fibers of ${\it Cl}_\ast M$.  Here ${\bf e}_I\doteq{\bf e}_{\alpha_1}{\bf e}_{\alpha_2}\cdots{\bf e}_{\alpha_k}$ for the multi--index $I=\alpha_1\alpha_2\cdots\alpha_k$.  We denote the length of $I$ by $|I|$; i.e., $|I|=k$.
 A field $\psi$ is a section $M\rightarrow{\it Cl}_\ast M$, and we write $\psi=\psi^I{\bf e}_I$, where $I=\emptyset,0,1,\dots,0123$.

The gamma matrix $\gamma_\mu$ is defined as left multiplication by the tangent vector ${\bf e}_\mu$.  That is, $\gamma_\mu\psi\doteq{\bf e}_{\mu}\psi$.  Moreover, we set $\gamma^\alpha\doteq g^{\alpha\beta}\gamma_\beta$.

The spacetime metric $g$ is extended to a metric $\hat{g}$ on ${\it Cl}_\ast M$ by the rule $\hat{g}(\psi,\phi)\doteq-\tfrac{1}{2}\langle\psi^\dagger\phi+\phi^\dagger\psi\rangle_\emptyset$.  Here $\psi^\dagger$ is the linear extension of ${\bf e}_{\alpha_1\alpha_2\cdots\alpha_k}^\dagger\doteq(-1)^k{\bf e}_{\alpha_k\cdots\alpha_2\cdots\alpha_1}$, and $\langle\psi\rangle_\emptyset$ is linear projection onto the ${\bf e}_\emptyset$ component of $\psi$.  One shows that $\gamma_\alpha^T\hat{g}+\hat{g}\gamma_\alpha=0$.

A transformation $A$ on the tangent bundle of $M$ extends to a transformation $\hat{A}$ on ${\it Cl}_\ast M$ via the rule $\hat{A}{\bf e}_{\alpha_1\alpha_2\cdots\alpha_k}\doteq(A{\bf e}_{\alpha_1})(A{\bf e}_{\alpha_2})\cdots(A{\bf e}_{\alpha_k})$.  In particular, suppose that $B$ is a change of basis for the tangent bundle of $M$: ${\bf e}_\alpha'=B^{-1}{\bf e}_\alpha$.  This extends to an action $\hat{B}$ on ${\it Cl}_\ast M$.  Under the extended change of basis, the gamma matrices transform as $\gamma_\alpha'=(B^{-1})_\alpha^\beta\hat{B}\gamma_\beta\hat{B}^{-1}$ and $\gamma'^\alpha=B_\beta^\alpha\hat{B}\gamma^\beta\hat{B}^{-1}$.  Whereas the extended metric transforms as $\hat{g}'=\hat{B}^{-T}\hat{g}\hat{B}^{-1}$.

The transformation extension rule in the previous paragraph applies to Lie groups.  That is, if $G$ is a Lie group acting locally on the tangent bundle, then we can extend the action of $G$ to ${\it Cl}_\ast M$.  On the other hand, the Lie algebra ${\mathfrak g}$ of $G$ extends to a Lie algebra action via the rule $\hat{a}{\bf e}_{\alpha_1\cdots\alpha_k}=\sum_{i=1}^k{\bf e}_{\alpha_1}\cdots{\bf e}_{\alpha_{i-1}}(a{\bf e}_{\alpha_i}){\bf e}_{\alpha_{i+1}}\dots{\bf e}_{\alpha_k}$ for all $a\in{\mathfrak g}$.

%%%%%%%%%%%%%%%%%%%%%%%%%%%%%%%%%%%%%%%%%%%%%%%%%%%%%%%%%%%%%%%%%
\subsection{Allowed forces}

We use minimal coupling to model forces.  That is, we assume there is a connection $\nabla$ on ${\it Cl}_\ast M$, where $\nabla_\alpha\psi=\partial_\alpha\psi+C_\alpha\psi$ for some collection of four $16\times 16$ matrices $C_\alpha$.  However, we need to ensure that we obtain the Dirac equation
$$D\psi+\mu\psi=0,
  \quad\text{where}\quad
  D\psi\doteq\gamma^\alpha\nabla_\alpha\psi,
$$
by varying the (partial) Lagrangian density
\begin{equation}\label{eq:diraclagrangian}
  L_K=(\psi^T\hat{g}D\psi+\mu\psi^T\hat{g}\psi)\,\omega,\quad
  \omega\doteq\sqrt{-\det(g)}
\end{equation}
with respect to $\psi$.  As observed in \cite{me}, the conditions
\begin{equation}\label{eq:fullcompat}
  C_\alpha^T\hat{g}+\hat{g}C_\alpha=\partial_\alpha\hat{g}
  \quad\text{and}\quad
  [\gamma^\alpha,C_\beta]
  =\partial_\beta\gamma^\alpha+\Gamma_{\beta\epsilon}^\alpha\gamma^\epsilon,
\end{equation}
where $\Gamma_{\beta\epsilon}^\alpha$ are the Christoffel symbols for the metric connection on $M$, are sufficient to guarantee this.

The metric connection on $M$ extends to a connection on ${\it Cl}_\ast M$ via the Leibniz rule.  Moreover, the extended metric connection matrices $\hat{\Gamma}_\alpha$ satisfy both conditions in equation \eqref{eq:fullcompat}.  So to incorporate forces other than gravity, we look for connection matrices of the form
$$C_\alpha=\hat\Gamma_\alpha+\theta_\alpha,
  \quad\text{with}
$$
\begin{equation}\label{eq:physcon}
  \theta_\alpha^T\hat{g}+\hat{g}\theta_\alpha=0
  \quad\text{and}\quad
  [\gamma^\alpha,\theta_\beta]=0.
\end{equation}
We call any collection $\theta_\alpha$ of $16\times 16$ matrices that satisfy \eqref{eq:physcon} a {\bf force tensor}.  We write $\theta_\ast$ to denote the collection $\{\theta_0,\theta_1,\theta_2,\theta_3\}$.

By general principles, under a local change of basis $B$ for the tangent bundle of $M$, the connection matrices $C_\alpha$ transform as
\begin{equation}\label{eq:xconn}
  C_\alpha'=(B^{-1})_\alpha^\beta(-\partial_\beta\hat{B}
                                  +\hat{B}C_\beta)\hat{B}^{-1}.
\end{equation}
The extended metric connection $\hat\Gamma_\alpha$ will satisfy this equation, so a force tensor must satisfy the transformation rule
\begin{equation}\label{eq:transform}
  \theta_\alpha'=(B^{-1})_\alpha^\beta\hat{B}\theta_\beta\hat{B}^{-1}
\end{equation}
Moreover, a force tensor remains a force tensor under a transformation.  That is, the equations in \eqref{eq:physcon} are satisfied by the transformed force tensor: $\theta_\alpha'^T\hat{g}'+\hat{g}'\theta_\alpha'=0$ and $[\gamma'^\alpha,\theta_\beta']=0$.

We remark that a force tensor is, in general, not the matrix of an actual connection.  It does not transform as a connection matrix is required to, equation \eqref{eq:xconn}.  However in the case of Minkowski space ${\amsmathbb M}$, where the extended metric connection is trivial $\hat\Gamma_\alpha=0$, a force tensor is indeed the matrix of a connection on the Clifford algebra ${\it Cl}({\amsmathbb M})$.

%%%%%%%%%%%%%%%%%%%%%%%%%%%%%%%%%%%%%%%%%%%%%%%%%%%%%%%%%%%%%%%%%
\section{The space of force tensors}\label{sec:space}

The set $V(g)$ of all force tensors is necessarily a (real) vector space.  We will determine a basis in the case of the Minkowski metric $g=\eta$, where
$$\eta\doteq{\rm diag}(-1,1,1,1).$$
Note that $\hat\eta$ is also diagonal.  The following fact, which we verify at the end of this section, facilitates the computation.

\begin{lemma}
Let $N$ be a $16\times 16$ matrix acting on ${\it Cl}({\amsmathbb M})$ as a vector space.  Then $N$ commutes with gamma matrices if and only if for all multi--indices $I$ we have $N{\bf e}_I={\bf e}_I\cdot N{\bf e}_\emptyset$.  Moreover for such $N$, $N^T\hat\eta+\hat\eta N=0$ if and only if $N{\bf e}_\emptyset$ lies in the subspace of ${\it Cl}({\amsmathbb M})$ spanned by the basis vectors ${\bf e}_I$ with $|I|=1,2$.
\end{lemma}

\noindent
Observe that from the first statement, any matrix $N$ that commutes with gamma matrices is completely determined by its effect on ${\bf e}_\emptyset$.

Let us introduce the following notation.  For a multi--index $I$, $\basisOne{I}$ denotes the $16\times 16$ matrix that commutes with gamma matrices and such that $\basisOne{I}{\bf e}_\emptyset={\bf e}_I$.  Moreover, we let $\basis{\beta}{I}_\ast$ denote the collection of matrices with
$$\basis{\beta}{I}_\alpha=0
  \,\,\text{if $\alpha\neq\beta$},
  \quad\text{and}\quad
  \basis{\beta}{I}_\alpha=\basisOne{I}
  \,\,\text{if $\alpha=\beta$}.
$$
I.e., $\basis{\beta}{I}_\alpha=\delta_{\alpha\beta}\basisOne{I}$.  It should be pointed out that even though $\basisOne{\alpha}{\bf e}_\emptyset={\bf e}_\alpha$ and $\gamma_\alpha{\bf e}_\emptyset={\bf e}_\alpha$, necessarily $\basisOne{\alpha}\neq\gamma_\alpha$.

From the above fact, we see that the space of force tensors $V(\eta)$ is a real vector space with basis given by
\begin{equation}\label{eq:basis}
  \basis{\beta}{I}_\ast
  \quad\text{with}\quad
  I=0,1,2,3,01,02,03,12,13,23
  \,\,\text{and}\,\,
  \beta=0,1,2,3.
\end{equation}
In particular, $V(\eta)$ has dimension $40$.

In the remainder of this section, we verify the previously claimed properties of the matrix $16\times 16$ matrix $N$.  The first statement is readily seen from the definition of the gamma matrices.  For the second statement, let $I=\alpha_1\cdots\alpha_k$ be a multi--index.  Set $\gamma_I\doteq\gamma_{\alpha_1}\cdots\gamma_{\alpha_k}$, and $\gamma_I^\dagger\doteq(-1)^k\gamma_{\alpha_k}\cdots\gamma_{\alpha_1}$, so that $\gamma_I{\bf e}_\emptyset={\bf e}_I$ and $\gamma_I^\dagger{\bf e}_\emptyset={\bf e}_I^\dagger$.  Moreover, $\hat\eta\gamma_I^\dagger=\gamma_I^T\hat\eta$.  Viewing ${\bf e}_I$ as a column vector, we have
$${\bf e}_\emptyset^T\hat\eta N{\bf e}_I
  ={\bf e}_\emptyset^T\hat\eta N\gamma_I{\bf e}_\emptyset
  ={\bf e}_\emptyset^T\hat\eta\gamma_I N{\bf e}_\emptyset.
$$
The symmetry of $\hat\eta$ implies that the quantity on the right is equal to
$${\bf e}_\emptyset^TN^T\gamma_I^T\hat\eta{\bf e}_\emptyset
  ={\bf e}_\emptyset^TN^T\hat\eta\gamma_I^\dagger{\bf e}_\emptyset
  ={\bf e}_\emptyset^TN^T\hat\eta{\bf e}_I^\dagger.
$$
It follows that ($\star$) ${\bf e}_\emptyset^T(\hat\eta N+N^T\hat\eta){\bf e}_I={\bf e}_\emptyset^TN^T\hat\eta({\bf e}_I^\dagger+{\bf e}_I)$ for any multi--index $I$.  Observe that ${\bf e}_I+{\bf e}_I^\dagger=0$ if $|I|=1,2$ and is equal to $2{\bf e}_I$ if $|I|=0,3,4$.  Therefore if $N^T\hat\eta+\hat\eta N=0$, ($\star$) and the fact that $\hat\eta$ is diagonal imply that $N{\bf e}_\emptyset$ can only lie in the subspace spanned by those ${\bf e}_I$ with $|I|=1,2$.

Conversely, let $J$ and $K$ be any multi--indices.  Now ${\bf e}_J^\dagger{\bf e}_K$ reduces to a multiple of a basis vector, say $r{\bf e}_I$.  So,
$${\bf e}_J^TN^T\hat\eta{\bf e}_K
  ={\bf e}_\emptyset^T\gamma_J^TN^T\hat\eta{\bf e}_K
  ={\bf e}_\emptyset^TN^T\gamma_J^T\hat\eta{\bf e}_K
  ={\bf e}_\emptyset^TN^T\hat\eta\gamma_J^\dagger{\bf e}_K
  =r{\bf e}_\emptyset^TN^T\hat\eta{\bf e}_I.
$$
Similarly, one computes ${\bf e}_J^T\hat\eta N{\bf e}_K=r{\bf e}_\emptyset^TN^T\hat\eta{\bf e}_I^\dagger$, so that ${\bf e}_J^T(N^T\hat\eta+\hat\eta N){\bf e}_K=r{\bf e}_\emptyset^TN^T\hat\eta({\bf e}_I+{\bf e}_I^\dagger)$.  Identity $(\star)$ thus implies that $N^T\hat\eta+\hat\eta N=0$ if $N{\bf e}_\emptyset$ lies in the stated subspace.

%%%%%%%%%%%%%%%%%%%%%%%%%%%%%%%%%%%%%%%%%%%%%%%%%%%%%%%%%%%%%%%%%
\section{Irreducible connections}\label{sec:irred}

A Lorentz transformation $\Lambda$ defines a change of basis for the tangent bundle of $M$.  We can extend this to a change of basis $\hat\Lambda$ for ${\it Cl}_\ast M$.  In this way the Lorentz group $O(g)$ acts on the space of force tensors $V(g)$.  Indeed, from equation \eqref{eq:transform}, 
$$(\Lambda\cdot\theta)_\alpha
  \doteq(\Lambda^{-1})_\alpha^\beta\hat{\Lambda}\theta_\beta\hat{\Lambda}^{-1}
$$
for any force tensor $\theta_\ast$.  The corresponding Lorentz algebra action is then
\begin{equation}\label{eq:action}
  (L\cdot\theta)_\alpha
  =-L_\alpha^\beta\theta_\beta
   +\hat{L}\theta_\alpha -\theta_\alpha\hat{L}
  =[\hat{L},\theta_\alpha]-L_\alpha^\beta \theta_\beta
\end{equation}
for $L\in so(g)$.

We say that a force tensor $\theta_\ast$ is {\bf irreducible} if its orbit under the $O(g)$ action, or equivalently under the $so(g)$ action, is an irreducible real representation of $O(g)$.  The irreducible sub--representations of $V(g)$ may be computed using standard Lie algebra techniques, such as in \cite{Fulton}.  We will present the results for the case $g=\eta$.

%::::::::::::::::::::::::::::::::::::::::::::::::::::::::::::::::
\subsection{Summary of fundamental computations}

We explicitly compute how $so(\eta)$ affects each of the basic force tensors $\basis{\beta}{I}_\ast$ in equation \eqref{eq:basis}.  As a vector space, $so(\eta)$ is six--dimensional, and we may take as basis the six $4\times 4$ matrices
$$s_{01},\,s_{02},\,s_{03},\,a_{12},\,a_{13},\,a_{23}$$
$$s_{0k}\doteq E_{0k}+E_{k0}
  \quad\text{and}\quad
  a_{jk}\doteq E_{jk}-E_{kj}
$$
where $E_{\alpha\beta}$ is the $4\times 4$ matrix whose $(\alpha,\beta)$--entry is unity, and all other entries are zero.  In particular, we have
\begin{equation}\label{eq:rep1}
  s_{0k}{\bf e}_0={\bf e}_k,\quad
  s_{0k}{\bf e}_k={\bf e}_0,\quad
  a_{jk}{\bf e}_j=-{\bf e}_k,\quad
  a_{jk}{\bf e}_k={\bf e}_j
\end{equation}
and $s_{0k}{\bf e}_l=0$ and $a_{jk}{\bf e}_l=0$ in all other cases.

The Lie algebra action of $so(\eta)$ on $\amsmathbb{M}$ extends to a Lie algebra action on ${\it Cl}(\amsmathbb{M})$.  For example,
$$\hat{s}_{03}{\bf e}_{02}
  =(s_{03}{\bf e}_0)({\bf e}_2)
   +({\bf e}_0)(s_{03}{\bf e}_2)
  =({\bf e}_3)({\bf e}_2)+({\bf e}_0)(0)
  =-{\bf e}_2{\bf e}_3
  =-{\bf e}_{23}
$$
Now from equation \eqref{eq:action}, the action of $L\in so(\eta)$ on $\basis{\beta}{I}$ is
$$(L\cdot\basis{\beta}{I})_\alpha
  =[\hat{L},\basis{\beta}{I}_\alpha]
   -L_\alpha^\nu\basis{\beta}{I}_\nu
  =\delta_\alpha^\beta[\hat{L},\basisOne{I}]
   -L_\alpha^\beta\basisOne{I}
$$
Figure \ref{table:rep0} gives a listing of values for $[\hat{L},\basisOne{I}]$ for our chosen basis matrices $L$ of $so(\eta)$.  For instance,
\begin{align*}
  \hat{s}_{03}\basisOne{02}{\bf e}_J
  &=\hat{s}_{03}{\bf e}_J{\bf e}_{02}
   =(\hat{s}_{03}{\bf e}_J)({\bf e}_{02})
    +({\bf e}_J)(\hat{s}_{03}{\bf e}_{02})\\
  &=\basisOne{02}\hat{s}_{03}{\bf e}_J-{\bf e}_J{\bf e}_{23}
   =\basisOne{02}\hat{s}_{03}{\bf e}_J-\basisOne{23}{\bf e}_J
\end{align*}
So that $[\hat{s}_{03},\basisOne{02}]=-\basisOne{23}$.  Figure \ref{table:rep0} and equation \eqref{eq:rep1} can then be used to compute the effect of the $so(\eta)$ action on the basis elements of $V(\eta)$.  E.g.,
\begin{align*}
  (a_{23}\cdot\basis{2}{12})_0
    &=0-(a_{23})_0^2\basisOne{12}=0\\
  (a_{23}\cdot\basis{2}{12})_1
    &=0-(a_{23})_1^2\basisOne{12}=0\\
  (a_{23}\cdot\basis{2}{12})_2
  &=[\hat{a}_{23},\basisOne{12}]-(a_{23})_2^2\basisOne{12}
   =-\basisOne{13}-0=-\basisOne{13}\\
  (a_{23}\cdot\basis{2}{12})_3
  &=0-(a_{23})_3^2\basisOne{12}
   =-\basisOne{12}
\end{align*}
Thus, $(a_{23}\cdot\basisOne{12})_\ast=(0,0,-\basisOne{13},-\basisOne{12})=-\basis{2}{13}_\ast-\basis{3}{12}_\ast$.

\begin{figure}
\begin{center}
\begin{tabular}{c|cccccc}
  $\basisOne{I}\backslash L$ & $s_{01}$ & $s_{02}$ & $s_{03}$
    & $a_{12}$ & $a_{13}$ & $a_{23}$\\\hline
  $\basisOne{0}$ & $\basisOne{1}$ & $\basisOne{2}$ & $\basisOne{3}$
    & $0$ & $0$ & $0$\\
  $\basisOne{1}$ & $\basisOne{0}$ & $0$ & $0$
    & $-\basisOne{2}$ & $-\basisOne{3}$ & $0$\\
  $\basisOne{2}$ & $0$ & $\basisOne{0}$ & $0$
    & $\basisOne{1}$ & $0$ & $-\basisOne{3}$\\
  $\basisOne{3}$ & $0$ & $0$ & $\basisOne{0}$
    & $0$ & $\basisOne{1}$ & $\basisOne{2}$\\
  $\basisOne{01}$ & $0$ & $-\basisOne{12}$ & $-\basisOne{13}$
    & $-\basisOne{02}$ & $-\basisOne{03}$ & $0$\\
  $\basisOne{02}$ & $\basisOne{12}$ & $0$ & $-\basisOne{23}$
    & $\basisOne{01}$ & $0$ & $-\basisOne{03}$\\
  $\basisOne{03}$ & $\basisOne{13}$ & $\basisOne{23}$ & $0$
    & $0$ & $\basisOne{01}$ & $\basisOne{02}$\\
  $\basisOne{12}$ & $\basisOne{02}$ & $-\basisOne{01}$ & $0$
    & $0$ & $\basisOne{23}$ & $-\basisOne{13}$\\
  $\basisOne{13}$ & $\basisOne{03}$ & $0$ & $-\basisOne{01}$
    & $-\basisOne{23}$ & $0$ & $\basisOne{12}$\\
  $\basisOne{23}$ & $0$ & $\basisOne{03}$ & $-\basisOne{02}$
    & $\basisOne{13}$ & $-\basisOne{12}$ & $0$\\
\end{tabular}
\end{center}
\caption{Table of values for $[\hat{L},\basisOne{I}]$.}\label{table:rep0}
\end{figure}

%::::::::::::::::::::::::::::::::::::::::::::::::::::::::::::::::
\subsection{Irreducible summands: overview}

Viewing the space of force tensors $V(g)$ as a Lie algebra representation of $so(g)$, we decompose it into irreducible summands.  Schematically, the decomposition is
\begin{equation}\label{eq:summands}
  V(g)=\mathbb{1}\oplus\mathbb{4}\oplus\mathbb{4}'
       \oplus\mathbb{6}\oplus\mathbb{9}\oplus{\mathbb{1}\!\mathbb{6}}
\end{equation}
where each summand has the indicated dimension.  The decomposition is over the reals.  Further decomposition of the summands $\mathbb{6}$ and $\mathbb{1}\!\mathbb{6}$ can be achieved over the complex numbers.  Indeed when complex coefficients are allowed, each of these two summands decomposes into two conjugate summands of complex dimensions equal to half the real dimension: $\mathbb{6}=\mathbb{3}\oplus\bar{\mathbb{3}}$ and $\mathbb{1}\!\mathbb{6}=\mathbb{8}\oplus\bar{\mathbb{8}}$.  The 4--dimensional summands $\mathbb{4}$ and $\mathbb{4}'$ are (real) isomorphic.  In fact, we will see that $\mathbb{4}$ and $\mathbb{4}'$ are both isomorphic to the standard 4--vector representation of the Lorentz group.  In the classification of representations of the Lorentz group, this representation is denoted $(\tfrac{1}{2},\tfrac{1}{2})$.  The summand $\mathbb{6}$ is isomorphic to $(1,0)\oplus(0,1)$, $\mathbb{9}$ isomorphic to $(1,1)$, and $\mathbb{1}\!\mathbb{6}$ isomorphic to $(\tfrac{3}{2},\tfrac{1}{2})\oplus(\tfrac{1}{2},\tfrac{3}{2})$.

%::::::::::::::::::::::::::::::::::::::::::::::::::::::::::::::::
\subsection{One--dimensional connection}\label{sec:con1}

The summand $\mathbb{1}$ is the one--dimensional subspace spanned by the single force tensor
$$u=\basis{0}{0}_\ast+\basis{1}{1}_\ast+\basis{2}{2}_\ast+\basis{3}{3}_\ast.
$$
Indeed, one computes that $L\cdot u=0$ for all generators $L=s_{01}$, $s_{02}$, $s_{03}$, $a_{12}$, $a_{13}$, $a_{23}$ of $so(\eta)$, and hence for all $L\in so(\eta)$.  That is, $so(\eta)$ acts trivially on $u$.  Any force tensor in $\mathbb{1}$ can thus be written in the form
\begin{equation}\label{eq:1-connection}
  U_\ast=\phi u
\end{equation}
for a scalar field $\phi$.  The field $\phi$ is necessarily unaffected by a change of basis; that is, its transformation rule is $\phi'=\phi$.  The individual component matrices of $U_\ast$ are
$$U_0=\phi\,\basisOne{0},\quad
  U_1=\phi\,\basisOne{1},\quad
  U_2=\phi\,\basisOne{2},\quad
  U_3=\phi\,\basisOne{3}
$$

%::::::::::::::::::::::::::::::::::::::::::::::::::::::::::::::::
\subsection{Four--dimensional connections}\label{ssec:4dcon}

There two four--dimensional summands in \eqref{eq:summands}.  We will see that the two summands are isomorphic.  As a consequence of this, if we set $V\doteq\mathbb{4}\oplus\mathbb{4}'$, then the decomposition of $V$ into irreducible summands is not unique: we can find irreducible subspaces $V_1$, $V_2$ of $V$ such that $V=V_1\oplus V_2$, but $V_1$ is equal to neither $\mathbb{4}$ nor $\mathbb{4}'$.  For the moment, we will choose summands that are comparatively easily to write down.  Specifically, set
$$V_1\doteq{\amsmathbb R}\{v_0,v_1,v_2,v_3\}
  \quad\text{and}\quad
  V_2\doteq{\amsmathbb R}\{v_0',v_1',v_2',v_3'\}$$
where
\begin{align*}
  v_0&\doteq -\basis{1}{01}_\ast-\basis{2}{02}_\ast-\basis{3}{03}_\ast
    & v_0'&\doteq -\basis{1}{23}_\ast+\basis{2}{13}_\ast-\basis{3}{12}_\ast\\
  v_1&\doteq\basis{0}{01}_\ast-\basis{2}{12}_\ast-\basis{3}{13}_\ast
    & v_1'&\doteq\basis{0}{23}_\ast+\basis{2}{03}_\ast-\basis{3}{02}_\ast\\
  v_2&\doteq\basis{0}{02}_\ast+\basis{1}{12}_\ast-\basis{3}{23}_\ast
    & v_2'&\doteq -\basis{0}{13}_\ast-\basis{1}{03}_\ast+\basis{3}{01}_\ast\\
  v_3&\doteq\basis{0}{03}_\ast+\basis{1}{13}_\ast+\basis{2}{23}_\ast
    & v_3'&\doteq\basis{0}{12}_\ast+\basis{1}{02}_\ast-\basis{2}{01}_\ast
\end{align*}
Observe that as matrices, $\basisOne{0123}\basisOne{01}=\basisOne{23}$.  In this way, we can write $v_\alpha'=Jv_\alpha$, so that $V_2=JV_1$, where $J\doteq\basisOne{0123}$.

In general, for any angle $\zeta$ we obtain an irreducible summand of $V$ by choosing basis vectors
$$\cos{\zeta}\,v_\alpha+\sin{\zeta}\,v_\alpha'
  =\cos{\zeta}\,v_\alpha+\sin{\zeta}\,Jv_\alpha
  =e^{\zeta J}v_\alpha
$$
where $e^{\zeta J}=\cos\zeta\,I+\sin\zeta\,J$ is the matrix exponential (note that $J^2=-I$).  Let us set
\begin{equation}\label{eq:4-zeta}
  \mathbb{4}_\zeta
  \doteq e^{\zeta J}V_1
  ={\amsmathbb R}\{e^{\zeta J}v_0,e^{\zeta J}v_1,
                   e^{\zeta J}v_2,e^{\zeta J}v_3\}.
\end{equation}
For different values of $\zeta$, we obtain different summands of $V$.  Moreover for $\zeta,\zeta'\in[0,\pi)$ with $\zeta\neq\zeta'$, $V=\mathbb{4}_\zeta\oplus\mathbb{4}_{\zeta'}$.

One computes the extended Lie algebra action of $so(\eta)$ on the basis vectors of $V_1$ to be
$$\hat{s}_{0k}v_0=v_k,\quad
  \hat{s}_{0k}v_k=v_0,\quad
  \hat{a}_{jk}v_j=-v_k,\quad
  \hat{a}_{jk}v_k=v_j
$$
for $j,k=1,2,3$ and $j\neq k$ (with all other actions trivial).  Moreover, the matrix $J$ commutes with the extended Lie algebra action, so that the basis vectors of $\mathbb{4}_\zeta$ in \eqref{eq:4-zeta} transform in the exact same way as those of $V_1$.  It follows that $\mathbb{4}_\zeta$ is indeed stable under the $so(\eta)$ action, and that all summands of this form are isomorphic.  Furthermore, the basis vectors of $\mathbb{4}_\zeta$ transform exactly like the basis vectors ${\bf e}_\alpha$ of ${\amsmathbb M}$ under Lorentz transformations.  Therefore, any force tensor in $\mathbb{4}_\zeta$ can be identified with a 4--vector field.  I.e., we have the force tensor
\begin{equation}\label{eq:4-connection}
  F_\ast=A^\alpha e^{\zeta J}\,v_\alpha
\end{equation}
for any 4--vector field $A^\alpha$.  Explicitly, we have
\begin{align*}
  F_0&=e^{\zeta J}( A^1\basisOne{01}
                   +A^2\basisOne{02}
                   +A^3\basisOne{03})\\
  F_1&=e^{\zeta J}(-A^0(\basisOne{01}
                   +A^2\basisOne{12}
                   +A^3\basisOne{13})\\
  F_2&=e^{\zeta J}(-A^0\basisOne{02}
                   -A^1\basisOne{12}
                   +A^3\basisOne{23})\\
  F_3&=e^{\zeta J}(-A^0\basisOne{03}
                   -A^1\basisOne{13}
                   -A^2\basisOne{23})
\end{align*}

%::::::::::::::::::::::::::::::::::::::::::::::::::::::::::::::::
\subsection{Six--dimensional connections}

The summand $\mathbb{6}$ in \eqref{eq:summands} is spanned by the force tensors
\begin{equation}\label{eq:6-connection-basis}
\begin{aligned}
    v_{01} &\doteq\basis{2}{3}_\ast-\basis{3}{2}_\ast
  & v_{02} &\doteq\basis{3}{1}_\ast-\basis{1}{3}_\ast\\
    v_{03} &\doteq\basis{1}{2}_\ast-\basis{2}{1}_\ast
  & v_{12} &\doteq\basis{3}{0}_\ast+\basis{0}{3}_\ast\\
    v_{13} &\doteq -\basis{2}{0}_\ast-\basis{0}{2}_\ast
  & v_{23} &\doteq\basis{1}{0}_\ast+\basis{0}{1}_\ast
\end{aligned}
\end{equation}
The action of $so(\eta)$ on these basic force tensors is given in table in figure \ref{table:so-6}.  This verifies that $\mathbb{6}$ is indeed stable under the $so(\eta)$ action.  In fact, this is exactly the same action as that of $so(\eta)$ on $\Lambda^2{\amsmathbb M}$, the vector space spanned by the 2--forms ${\bf e}_\alpha\wedge{\bf e}_\beta\doteq{\bf e}_\alpha\otimes{\bf e}_\beta-{\bf e}_\beta\otimes{\bf e}_\alpha$.  Hence we may identify any connection from $\mathbb{6}$ with an anti--symmetric tensor field $W^{\alpha\beta}$.  That is, we may write
\begin{equation}\label{eq:6-connection}
  H_\ast=W^{\alpha\beta}v_{\alpha\beta}
\end{equation}
for some anti--symmetric tensor field $W^{\alpha\beta}$, provided we {\em define} the basic connections $v_{\alpha\beta}$ to satisfy $v_{\alpha\beta}=0$ if $\alpha\geq\beta$.  Explicitly,
\begin{align*}
  H_0&=W^{12}\basisOne{3}-W^{13}\basisOne{2}+W^{23}\basisOne{1}\\
  H_1&=-W^{02}\basisOne{3}+W^{03}\basisOne{2}+W^{23}\basisOne{0}\\
  H_2&=W^{01}\basisOne{3}-W^{03}\basisOne{1}-W^{13}\basisOne{0}\\
  H_3&=-W^{01}\basisOne{2}+W^{02}\basisOne{1}+W^{12}\basisOne{0}
\end{align*}

\begin{figure}
\begin{center}
\begin{tabular}{c|cccccc}
           & $s_{01}$ & $s_{02}$ & $s_{03}$ & $a_{12}$ & $a_{13}$ & $a_{23}$\\\hline
  $v_{01}$ & $0$ & $-v_{12}$ & $-v_{13}$ & $-v_{02}$ & $-v_{03}$ & $0$\\
  $v_{02}$ & $v_{12}$ & $0$ & $-v_{23}$ & $v_{01}$ & $0$ & $-v_{03}$\\
  $v_{03}$ & $v_{13}$ & $v_{23}$ & $0$ & $0$ & $v_{01}$ & $v_{02}$\\
  $v_{12}$ & $v_{02}$ & $-v_{01}$ & $0$ & $0$ & $v_{23}$ & $-v_{13}$\\
  $v_{13}$ & $v_{03}$ & $0$ & $-v_{01}$ & $-v_{23}$ & $0$ & $v_{12}$\\
  $v_{23}$ & $0$ & $v_{03}$ & $-v_{02}$ & $v_{13}$ & $-v_{12}$ & $0$\\
\end{tabular}
\end{center}
\caption{The action of $so(\eta)$ on the basic force tensors of $\mathbb{6}$.}\label{table:so-6}
\end{figure}

%::::::::::::::::::::::::::::::::::::::::::::::::::::::::::::::::
\subsection{Nine--dimensional connections}

A basis for summand $\mathbb{9}$ in \eqref{eq:summands} is given by the following basic force tensors.
\begin{equation}\label{eq:9-connection-basis}
\begin{aligned}
    u_{00} &\doteq -\basis{0}{0}_\ast+\basis{3}{3}_\ast
  & u_{01} &\doteq  \basis{1}{0}_\ast-\basis{0}{1}_\ast\\
    u_{02} &\doteq  \basis{2}{0}_\ast-\basis{0}{2}_\ast
  & u_{03} &\doteq  \basis{3}{0}_\ast-\basis{0}{3}_\ast\\
    u_{11} &\doteq  \basis{1}{1}_\ast-\basis{3}{3}_\ast
  & u_{12} &\doteq  \basis{2}{1}_\ast+\basis{1}{2}_\ast\\
    u_{13} &\doteq  \basis{3}{1}_\ast+\basis{1}{3}_\ast
  & u_{22} &\doteq  \basis{2}{2}_\ast-\basis{3}{3}_\ast\\
    u_{23} &\doteq  \basis{3}{2}_\ast+\basis{2}{3}_\ast
\end{aligned}
\end{equation}
The action of $so(\eta)$ on these is summarized in the table in figure \ref{table:so-9}, from which we see that the above collection of force tensors is stable under the $so(\eta)$ action.

We show that the basic force tensors in \eqref{eq:9-connection-basis} form a basis for the space of symmetric traceless $4\times 4$ matrices on Minkowski space ${\amsmathbb M}$.  Indeed, suppose $S^{\alpha\beta}$ is such that $S^{\beta\alpha}=S^{\alpha\beta}$ and $S_\alpha^\alpha=\eta_{\alpha\beta}S^{\alpha\beta}=0$.  The latter condition is equivalent to $S^{33}=S^{00}-S^{11}-S^{22}$.  Therefore, we may write
\begin{align*}
  S^{\alpha\beta}{\bf e}_\alpha\otimes{\bf e}_\beta
  &=S^{00}({\bf e}_0\otimes{\bf e}_0+{\bf e}_3\otimes{\bf e}_3)
    +S^{01}({\bf e}_0\otimes{\bf e}_1+{\bf e}_1\otimes{\bf e}_0)\\
  &\quad\quad
    +S^{02}({\bf e}_0\otimes{\bf e}_2+{\bf e}_2\otimes{\bf e}_0)
    +S^{03}({\bf e}_0\otimes{\bf e}_3+{\bf e}_3\otimes{\bf e}_0)\\
  &\quad\quad
    +S^{11}({\bf e}_1\otimes{\bf e}_1-{\bf e}_3\otimes{\bf e}_3)
    +S^{12}({\bf e}_1\otimes{\bf e}_2+{\bf e}_2\otimes{\bf e}_1)\\
  &\quad\quad
    +S^{13}({\bf e}_1\otimes{\bf e}_3+{\bf e}_3\otimes{\bf e}_1)
    +S^{22}({\bf e}_2\otimes{\bf e}_2-{\bf e}_3\otimes{\bf e}_3)\\
  &\quad\quad
    +S^{23}({\bf e}_2\otimes{\bf e}_3+{\bf e}_3\otimes{\bf e}_2)
\end{align*}
which defines a basis for the space of symmetric traceless matrices.  One computes that the $so(\eta)$ action on this basis is exactly the same as the action on the corresponding basic force tensors in equation \eqref{eq:9-connection-basis}.

In sum, if we define $u_{\alpha\beta}=0$ for $\alpha>\beta$ and $u_{33}=0$, then a force tensor in $\mathbb{9}$ can be written in the form
$$N_\ast=S^{\alpha\beta}u_{\alpha\beta}$$
for some symmetric traceless tensor field $S^{\alpha\beta}$; i.e., such that $S^{\alpha\beta}=S^{\beta\alpha}$ and $S_\alpha^\alpha=0$.  Explicitly,
\begin{align*}
  N_0&=-S^{00}\basisOne{0}-S^{01}\basisOne{1}
       -S^{02}\basisOne{2}-S^{03}\basisOne{3}\\
  N_1&= S^{01}\basisOne{0}+S^{11}\basisOne{1}
       +S^{12}\basisOne{2}+S^{13}\basisOne{3}\\
  N_2&= S^{02}\basisOne{0}+S^{12}\basisOne{1}
       +S^{22}\basisOne{2}+S^{23}\basisOne{3}\\
  N_3&= S^{03}\basisOne{0}+S^{13}\basisOne{1}
       +S^{23}\basisOne{2}+S^{33}\basisOne{3}
\end{align*}

\begin{figure}
\begin{center}
\begin{tabular}{c|cccccc}
  & $s_{01}$ & $s_{02}$ & $s_{03}$
     & $a_{12}$ & $a_{13}$ & $a_{23}$\\\hline
  $u_{00}$ & $u_{01}$ & $u_{02}$ & $2u_{03}$
     & $0$ & $u_{13}$ & $u_{23}$\\
  $u_{01}$ & $2u_{00}+2u_{11}$ & $u_{12}$ & $u_{13}$
     & $-u_{02}$ & $-u_{03}$ & $0$\\
  $u_{02}$ & $u_{12}$ & $2u_{00}+2u_{22}$ & $u_{23}$
     & $u_{01}$ & $0$ & $-u_{03}$\\
  $u_{03}$ & $u_{13}$ & $u_{23}$ & $2u_{00}$
     & $0$ & $u_{01}$ & $u_{02}$\\
  $u_{11}$ & $u_{01}$ & $0$ & $-u_{03}$
     & $-u_{12}$ & $-2u_{13}$ & $-u_{23}$\\
  $u_{12}$ & $u_{02}$ & $u_{01}$ & $0$
     & $2u_{11}-2u_{22}$ & $-u_{23}$ & $-u_{13}$\\
  $u_{13}$ & $u_{03}$ & $0$ & $u_{01}$
     & $-u_{23}$ & $2u_{11}$ & $u_{12}$\\
  $u_{22}$ & $0$ & $u_{02}$ & $-u_{03}$
     & $u_{12}$ & $-u_{13}$ & $-2u_{23}$\\
  $u_{23}$ & $0$ & $u_{03}$ & $u_{02}$
     & $u_{13}$ & $u_{12}$ & $2u_{22}$
\end{tabular}
\end{center}
\caption{The action of $so(\eta)$ on the basic force tensors of $\mathbb{9}$.}\label{table:so-9}
\end{figure}

%::::::::::::::::::::::::::::::::::::::::::::::::::::::::::::::::
\subsection{Sixteen--dimensional connections}

The sixteen--dimensional summand $\mathbb{1}\!\mathbb{6}$ in equation \ref{eq:summands} is spanned by the following force tensors:
\begin{equation}\label{eq:16-connection-basis}
\begin{aligned}
    u_{102}&\doteq\basis{1}{02}_\ast-\basis{0}{12}_\ast
  & u_{103}&\doteq\basis{1}{03}_\ast-\basis{0}{13}_\ast\\
    u_{112}&\doteq\basis{1}{12}_\ast-\basis{0}{02}_\ast
  & u_{113}&\doteq\basis{1}{13}_\ast-\basis{0}{03}_\ast\\
    u_{201}&\doteq\basis{2}{01}_\ast+\basis{0}{12}_\ast
  & u_{202}&\doteq\basis{2}{02}_\ast+\basis{1}{01}_\ast\\
    u_{203}&\doteq\basis{2}{03}_\ast-\basis{0}{23}_\ast
  & u_{212}&\doteq\basis{2}{12}_\ast+\basis{0}{01}_\ast\\
    u_{213}&\doteq\basis{2}{13}_\ast+\basis{1}{23}_\ast
  & u_{223}&\doteq\basis{2}{23}_\ast-\basis{0}{03}_\ast\\
    u_{301}&\doteq\basis{3}{01}_\ast+\basis{0}{13}_\ast
  & u_{302}&\doteq\basis{3}{02}_\ast+\basis{0}{23}_\ast\\
    u_{303}&\doteq\basis{3}{03}_\ast-\basis{1}{01}_\ast
  & u_{312}&\doteq\basis{3}{12}_\ast-\basis{1}{23}_\ast\\
    u_{313}&\doteq\basis{3}{13}_\ast+\basis{0}{01}_\ast
  & u_{323}&\doteq\basis{3}{23}_\ast+\basis{0}{02}_\ast
\end{aligned}
\end{equation}
The action of $so(\eta)$ on these force tensors is given by the table in figure \ref{table:so-16}.  We will identify the coefficients of a force tensor in $\mathbb{1}\!\mathbb{6}$ with a Fierz tensor.  Such tensors are described briefly in \cite{Arcos}.

Consider the subspace ${\amsmathbb F}$ of the tensor product space ${\amsmathbb M}\otimes{\amsmathbb M}\otimes{\amsmathbb M}$ consisting of tensors $F^{\alpha\rho\sigma}$ of the form
\begin{equation}\label{eq:fierz}
  F^{\alpha\rho\sigma}=-F^{\alpha\sigma\rho},
  \quad
  \eta_{\alpha\rho}F^{\alpha\rho\sigma}=0,
  \quad
  \epsilon_{\alpha\rho\sigma\tau}F^{\alpha\rho\sigma}=0.
\end{equation}
These three constraints allow us to write
\begin{align*}
  F^{\alpha\rho\sigma}
  {\bf e}_\alpha\otimes{\bf e}_\rho\otimes{\bf e}_\sigma
  =&F^{102}({\bf f}_{102}+{\bf f}_{012})
    +F^{103}({\bf f}_{103}+{\bf f}_{013})
    +F^{112}({\bf f}_{112}+{\bf f}_{002})\\
  &\mbox{}+F^{113}({\bf f}_{113}+{\bf f}_{003})
   +F^{201}({\bf f}_{201}-{\bf f}_{012})
   +F^{202}({\bf f}_{202}-{\bf f}_{101})\\
  &\mbox{}+F^{203}({\bf f}_{203}+{\bf f}_{023})
   +F^{212}({\bf f}_{212}-{\bf f}_{001})
   +F^{213}({\bf f}_{213}+{\bf f}_{123})\\
  &\mbox{}+F^{223}({\bf f}_{223}+{\bf f}_{003})
   +F^{301}({\bf f}_{301}-{\bf f}_{013})
   +F^{302}({\bf f}_{302}-{\bf f}_{023})\\
  &\mbox{}+F^{303}({\bf f}_{303}-{\bf f}_{101})
   +F^{312}({\bf f}_{312}-{\bf f}_{123})
   +F^{313}({\bf f}_{313}-{\bf f}_{001})\\
  &\mbox{}+F^{323}({\bf f}_{323}-{\bf f}_{002})
\end{align*}
where ${\bf f}_{\alpha\rho\sigma}\doteq{\bf e}_\alpha\otimes{\bf e}_\rho\otimes{\bf e}_\sigma-{\bf e}_\alpha\otimes{\bf e}_\sigma\otimes{\bf e}_\rho$.  One computes that the action of $so(\eta)$ on the basis of ${\amsmathbb F}$ thus defined is exactly as the table in figure \ref{table:so-16}.  That is, the basis element ${\bf f}_{102}+{\bf f}_{012}$ transforms in the same way as the basic force tensor $u_{102}$, ${\bf f}_{103}+{\bf f}_{013}$ transforms exactly as $u_{103}$, and so on.

It follows that a general force tensor in $\mathbb{1}\!\mathbb{6}$ can be expressed in the form
$$X_\ast=F^{\alpha\rho\sigma}u_{\alpha\rho\sigma},
$$
where $F^{\alpha\rho\sigma}$ satisfies equation \eqref{eq:fierz}, and provided we set $u_{\alpha\rho\sigma}=0$ for indices that do not match those of the stated basis elements.  E.g., $u_{001}=0$, $u_{231}=0$, et cetera.  Moreover, we have
{\small
\begin{align*}
  X_0&=-F^{001}\basisOne{01}
       -F^{002}\basisOne{02}
       -F^{003}\basisOne{03}
       -F^{012}\basisOne{12}
       -F^{013}\basisOne{13}
       -F^{023}\basisOne{23}\\
  X_1&= F^{101}\basisOne{01}
       +F^{102}\basisOne{02}
       +F^{103}\basisOne{03}
       +F^{112}\basisOne{12}
       +F^{113}\basisOne{13}
       +F^{123}\basisOne{23}\\
  X_2&= F^{201}\basisOne{01}
       +F^{202}\basisOne{02}
       +F^{203}\basisOne{03}
       +F^{212}\basisOne{12}
       +F^{213}\basisOne{13}
       +F^{223}\basisOne{23}\\
  X_3&= F^{301}\basisOne{01}
       +F^{302}\basisOne{02}
       +F^{303}\basisOne{03}
       +F^{312}\basisOne{12}
       +F^{313}\basisOne{13}
       +F^{323}\basisOne{23}
\end{align*}
}

\begin{figure}
\begin{center}
{\scriptsize
\begin{tabular}{c|cccccc}
        & $s_{01}$ & $s_{02}$ & $s_{03}$
          & $a_{12}$ & $a_{13}$ & $a_{23}$\\\hline
$u_{102}$ & $2u_{112}$ & $u_{212}$ & $u_{312}$
          & $-u_{202}$ & $-u_{302}$ & $-u_{103}$\\
$u_{103}$ & $2u_{113}$ & $u_{213}$ & $u_{313}$
          & $-u_{203}$ & $-u_{303}$ & $u_{102}$\\
$u_{112}$ & $2u_{102}$ & $u_{202}$ & $u_{302}$
          & $-u_{212}$ & $-u_{312}$ & $-u_{113}$\\
$u_{113}$ & $2u_{103}$ & $u_{203}$ & $u_{303}$
          & $-u_{213}$ & $-u_{313}$ & $u_{112}$\\
$u_{201}$ & $-u_{112}$ & $-2u_{212}$ & $-u_{312}-u_{213}$
          & $-u_{202}$ & $-u_{203}$ & $-u_{301}$\\
$u_{202}$ & $u_{212}$ & $u_{112}$ & $u_{113}-u_{223}$
          & $2u_{102}+2u_{201}$ & $u_{301}+u_{103}$ & $-u_{302}-u_{203}$\\
$u_{203}$ & $u_{213}$ & $2u_{223}$ & $u_{323}$
          & $u_{103}$ & $u_{201}$ & $u_{202}-u_{303}$\\
$u_{212}$ & $u_{202}$ & $-2u_{201}$ & $-u_{301}$
          & $u_{112}$ & $u_{223}$ & $-u_{213}-u_{312}$\\
$u_{213}$ & $u_{203}$ & $u_{103}$ & $-u_{102}-u_{201}$
          & $2u_{113}-2u_{223}$ & $-u_{112}-u_{323}$ & $u_{212}-u_{313}$\\
$u_{223}$ & $u_{103}$ & $2u_{203}$ & $u_{303}-u_{202}$
          & $u_{213}$ & $-u_{212}$ & $-u_{323}$\\
$u_{301}$ & $-u_{113}$ & $-u_{213}-u_{312}$ & $-2u_{313}$
          & $-u_{302}$ & $-u_{303}$ & $u_{201}$\\
$u_{302}$ & $u_{312}$ & $-u_{223}$ & $-2u_{323}$
          & $u_{301}$ & $u_{102}$ & $u_{202}-u_{303}$\\
$u_{303}$ & $u_{313}$ & $u_{323}+u_{112}$ & $u_{113}$
          & $u_{102}+u_{201}$ & $2u_{103}+2u_{301}$ & $u_{203}+u_{302}$\\
$u_{312}$ & $u_{302}$ & $-u_{103}-u_{301}$ & $u_{102}$
          & $u_{223}-u_{113}$ & $2u_{112}+2u_{323}$ & $u_{212}-u_{313}$\\
$u_{313}$ & $u_{303}$ & $-u_{201}$ & $-2u_{301}$
          & $-u_{323}$ & $u_{113}$ & $u_{213}+u_{312}$\\
$u_{323}$ & $-u_{102}$ & $u_{303}-u_{202}$ & $-2u_{302}$
          & $u_{313}$ & $-u_{312}$ & $u_{223}$
\end{tabular}
}
\end{center}
\caption{The action of $so(\eta)$ on the basic force tensors of $\mathbb{1}\!\mathbb{6}$.}\label{table:so-16}
\end{figure}

%%%%%%%%%%%%%%%%%%%%%%%%%%%%%%%%%%%%%%%%%%%%%%%%%%%%%%%%%%%%%%%%%
\section{Fundamental forces}\label{sec:force}

Each distinct irreducible summand in \eqref{eq:summands} corresponds to a distinct fundamental force.  That is, there are five fundamental forces (not including gravity) predicted by this model.  In this section, for each fundamental force we compute the the potential energy term under the assumption
$$L_V\doteq\omega\,{\it tr}(\Omega_{\alpha\beta}\Omega^{\alpha\beta}),$$
where $\Omega_{\alpha\beta}=\partial_\alpha C_\beta-\partial_\beta C_\alpha+C_\alpha C_\beta-C_\beta C_\alpha$ is the curvature matrix of the total connection $C_\alpha=\hat\Gamma_\alpha+\theta_\alpha$, with $\theta_\ast$ the force tensor.  We also compute the force field equations obtained by varying the Lagrangian density
\begin{equation}\label{eq:lagrangian}
  L=L_K+\tau L_V,
\end{equation}
with $L_K$ as in equation \eqref{eq:diraclagrangian} and $\tau$ a constant, with respect to the force field components.  Again we assume that $g=\eta$, the Minkowski metric, so that $C_\alpha=\theta_\alpha$.

With the exception of the scalar field, all computations were performed with the assistance of a symbolic algebra package.

%::::::::::::::::::::::::::::::::::::::::::::::::::::::::::::::::
\subsection{Scalar field}

The potential energy of the force tensor $U_\ast$ in equation \eqref{eq:1-connection}, associated with the one--dimensional irreducible summand $\mathbb{1}$, is found to be
$$L_V=96\,(\partial^\alpha\phi)(\partial_\alpha\phi)-768\,\phi^4$$
Variation of the Lagrangian density \eqref{eq:lagrangian} with respect to the scalar field $\phi$ then gives the field equation
\begin{equation}\label{eq:scalar}
  \partial_\alpha\partial^\alpha\phi+16\phi^3
  =\tfrac{1}{192\tau_1}\psi^T\hat\eta\gamma^\alpha\basisOne{\alpha}\psi
\end{equation}
for some constant $\tau_1$.

%::::::::::::::::::::::::::::::::::::::::::::::::::::::::::::::::
\subsection{Vector field}

The potential energy term in the Lagrangian for the force tensor $F_\ast$ on $\mathbb{4}_\zeta$ given in equation \eqref{eq:4-connection} is
\begin{align*}
  L_V=&-384(A_\alpha A^\alpha)^2\cos{4\zeta}
        +256(A^\alpha A_\alpha\partial_\beta A^\beta
             -A^\alpha A_\beta\partial_\alpha A^\beta)\cos{3\zeta}\\
     &-32[2(\partial_\alpha A^\beta)(\partial^\alpha A_\beta)
               +(\partial_\alpha A^\alpha)^2]\cos{2\zeta}
      +32\epsilon^{\alpha\beta\rho\sigma}
          (\partial_\alpha A_\beta)(\partial_\rho A_\sigma)\sin{2\zeta}
\end{align*}
Here $\epsilon^{\alpha\beta\rho\sigma}$ is the totally antisymmetric tensor of rank four.  The field equation for the four--vector field $A^\alpha$ is then
\begin{align*}
  &(\partial_\beta\partial^\beta A_\alpha
   +\partial_\alpha\partial_\beta A^\beta)\cos{2\zeta}
   -6(A_\beta\partial_\alpha A^\beta
       -A_\alpha\partial_\beta A^\beta)\cos{3\zeta}\\
  &\quad\quad
   -12(A_\beta A^\beta)A_\alpha\cos{4\zeta}
   =-\tfrac{1}{128\tau_4}\psi^T\hat\eta\gamma^\beta e^{\zeta J}(v_\alpha)_\beta
\end{align*}
where $v_\alpha$ is the basic force tensor from section \ref{ssec:4dcon}, and $(v_\alpha)_\beta$ is the $\beta$--component matrix of $v_\alpha$

%::::::::::::::::::::::::::::::::::::::::::::::::::::::::::::::::
\subsection{Antisymmetric tensor field}

In this case, the potential energy term of the Lagrangian for the connection $H_\ast$ in \eqref{eq:6-connection} is
\begin{align*}
  L_V
  &=-32(W^{\alpha\beta}W_{\alpha\beta})^2
      +4(\epsilon_{\alpha\beta\rho\sigma}W^{\alpha\beta}W^{\rho\sigma})^2\\
  &\quad\quad
      \mbox{}-16(\partial_\alpha W^{\rho\sigma})(\partial^\alpha W_{\rho\sigma})
      -32(\partial_\beta W^{\alpha\beta})(\partial^\sigma W_{\alpha\sigma})
\end{align*}
The field equation for the antisymmetric tensor field $W^{\alpha\beta}$ is
\begin{align*}
  &\partial_\mu\partial^\mu W_{\alpha\beta}
   -\partial_\alpha\partial^\sigma W_{\sigma\beta}
   +\partial_\beta\partial^\sigma W_{\sigma\alpha}
   -4(W^{\rho\sigma}W_{\rho\sigma})W_{\alpha\beta}\\
  &\quad\quad
  \mbox{}
  +\tfrac{1}{2}(\epsilon_{\mu\nu\rho\sigma}W^{\mu\nu}W^{\rho\sigma})
                  \epsilon_{\alpha\beta\kappa\lambda}W^{\kappa\lambda}
   =-\tfrac{1}{64\tau_6}\psi^T\hat\eta\gamma^\sigma
               (v_{\alpha\beta}-v_{\beta\alpha})_\sigma\psi
\end{align*}
where $(v_{\alpha\beta})_\sigma$ is the $\sigma$--component matrix of the basic force tensor $v_{\alpha\beta}$ in equation \eqref{eq:6-connection-basis}.

%::::::::::::::::::::::::::::::::::::::::::::::::::::::::::::::::
\subsection{Symmetric tensor field}

For a symmetric traceless tensor field $S^{\alpha\beta}$, the potential energy term for the Lagrangian is
\begin{align*}
  L_V&=-32(S^{\alpha\beta}S_{\alpha\beta})^2
       +256\,{\rm det}(S)
       +32(\partial^\alpha S^{\beta\rho})(\partial_\alpha S_{\beta\rho})
       -32(\partial^\alpha S^{\beta\rho})(\partial_\beta S_{\alpha\rho})
\end{align*}
where ${\rm det}(S)$ is the determinant of the field $S^{\alpha\beta}$ as a $4\times 4$ matrix.  To compute the variation of the total Lagrangian with respect to $S^{\alpha\beta}$, we need to take into the constraint $S^{33}=S^{00}-S^{11}-S^{22}$.  The result is
\begin{align*}
      \xi_{00}+\xi_{33}=\chi_{00},
  & & \xi_{01}=\chi_{01},
  & & \xi_{02}=\chi_{02},\\
      \xi_{03}=\chi_{03},
  & & \xi_{11}-\xi_{33}=\chi_{11},
  & & \xi_{12}=\chi_{12},\\
      \xi_{13}=\chi_{13},
  & & \xi_{22}-\xi_{33}=\chi_{22},
  & & \xi_{23}=\chi_{23}
\end{align*}
where
\begin{align*}
   \xi_{\alpha\beta}
   &\doteq\partial_\mu\partial^\mu S_{\alpha\beta}
      -\tfrac{1}{2}\partial_\alpha\partial^\mu S_{\mu\beta}
      -\tfrac{1}{2}\partial_\beta\partial^\mu S_{\mu\alpha}\\
   &\quad\quad\mbox{}
      +2(S^{\rho\sigma}S_{\rho\sigma})S_{\alpha\beta}
      -2\,[{\rm cof}(S)_{\alpha\beta}+{\rm cof}(S)_{\beta\alpha}]\\
   \chi_{\alpha\beta}
   &\doteq\tfrac{1}{128\tau_9}\psi^T\hat\eta\gamma^\rho
      (u_{\alpha\beta}+u_{\beta\alpha})_\rho\psi
\end{align*}
Here, ${\rm cof}(S)$ is the cofactor matrix of $S$, and $u_{\alpha\beta}$ are the basic force tensors in equation \eqref{eq:9-connection-basis}.

%::::::::::::::::::::::::::::::::::::::::::::::::::::::::::::::::
\subsection{Fierz tensor field}

In this case, the potential energy term of the Lagrangian is given by
\begin{align*}
  L_V&=-8(F^{\alpha\rho\sigma}F_{\alpha\rho\sigma})^2
       +2(\epsilon_{\rho\sigma\mu\nu}F^{\alpha\rho\sigma}
          {F_\alpha}^{\mu\nu})^2
       +16(\partial_\beta F^{\alpha\rho\sigma})
          (\partial_\alpha{F^{\beta}}_{\rho\sigma})\\
     &\quad
      -16(\partial_\beta F^{\alpha\rho\sigma})
         (\partial^\beta F_{\alpha\rho\sigma})
      +128(\partial_\beta F^{\alpha\rho\sigma})
          {F_{\alpha\rho}}^\mu
          {F^\beta}_{\sigma\mu}
\end{align*}
where $F^{\alpha\rho\sigma}$ is a Fierz tensor; i.e., a tensor that satisfies equation \eqref{eq:fierz}.  Constrained variation of the total Lagrangian yields
\begin{align*}
    \xi_{102}+\xi_{012}&=\chi_{102},
  & \xi_{103}+\xi_{013}&=\chi_{103},
  & \xi_{112}+\xi_{002}&=\chi_{112},
  & \xi_{113}+\xi_{003}&=\chi_{113},\\
    \xi_{201}-\xi_{012}&=\chi_{201},
  & \xi_{202}-\xi_{101}&=\chi_{202},
  & \xi_{203}+\xi_{023}&=\chi_{203},
  & \xi_{212}-\xi_{001}&=\chi_{212},\\
    \xi_{213}+\xi_{123}&=\chi_{213},
  & \xi_{223}+\xi_{003}&=\chi_{223},
  & \xi_{301}-\xi_{013}&=\chi_{301},
  & \xi_{302}-\xi_{023}&=\chi_{302},\\
    \xi_{303}-\xi_{101}&=\chi_{303},
  & \xi_{312}-\xi_{123}&=\chi_{312},
  & \xi_{313}-\xi_{001}&=\chi_{313},
  & \xi_{323}-\xi_{002}&=\chi_{323},
\end{align*}
where
\begin{align*}
  \xi_{\alpha\rho\sigma}
  &\doteq\partial_\beta\partial^\beta F_{\alpha\rho\sigma}
         -\partial_\alpha\partial_\beta{F^\beta}_{\rho\sigma}
         +4{F^{\beta\nu}}_\sigma\partial_\beta F_{\alpha\rho\nu}
         -4{F^{\beta\nu}}_\rho\partial_\beta F_{\alpha\sigma\nu}\\
  &\quad\mbox{}
         +2{F^{\kappa\lambda}}_\sigma\partial_\alpha F_{\kappa\lambda\rho}
         -2{F^{\kappa\lambda}}_\rho\partial_\alpha F_{\kappa\lambda\sigma}
         +2{F_{\alpha\sigma}}^\kappa\partial_\beta{F^\beta}_{\rho\kappa}
         -2{F_{\alpha\rho}}^\kappa\partial_\beta{F^\beta}_{\sigma\kappa}\\
  &\quad\mbox{}
         -(F^{\beta\mu\nu}F_{\beta\mu\nu})F_{\alpha\rho\sigma}
         +\tfrac{1}{4}(\epsilon_{\kappa\lambda\mu\nu}
                        F^{\beta\kappa\lambda}{F_\beta}^{\mu\nu})
                       \epsilon_{\rho\sigma\theta\phi}{F_\alpha}^{\theta\phi}\\
  \chi_{\alpha\rho\sigma}
  &\doteq -\tfrac{1}{64\tau_{16}}\psi^T\hat\gamma^\nu
           (u_{\alpha\rho\sigma}-u_{\alpha\sigma\rho})_\nu\psi
\end{align*}
Here, $u_{\alpha\rho\sigma}$ are the basic force tensors in equation \eqref{eq:16-connection-basis}.

%%%%%%%%%%%%%%%%%%%%%%%%%%%%%%%%%%%%%%%%%%%%%%%%%%%%%%%%%%%%%%%%%


\begin{thebibliography}{66}

\bibitem{Arcos} H. I. Arcos, C. S. O. Mayor, G. Ot\'{a}lora, and J. G. Pereira, {\it Spin--2 fields and helicity,} Found. Phys. (2012) 42: 1339--1349.

\bibitem{Fulton} William Fulton and Joe Harris, {\it Representation Theory: A First Course,} Graduate Texts in Mathematics, Springer, New York, 1991.

\bibitem{me} jason hanson, {\it Coupling the Dirac and Einstein equations through geometry,} arXiv:1903.11792.

\end{thebibliography}
\end{document}